\documentclass[aps,prd,preprint,showpacs]{revtex4}
\usepackage{amsmath,amssymb,amsthm,graphicx}
\usepackage[mathscr]{eucal}
\usepackage[colorlinks,linkcolor=blue,anchorcolor=blue,citecolor=green]{hyperref}
\usepackage[dvipsnames,usenames]{color}
\usepackage{graphicx}
\usepackage{subfigure}
\usepackage{amsmath}
\newcommand{\no}{\nonumber}
\newcommand{\be}{\begin{equation}}
\newcommand{\ee}{\end{equation}}
\newcommand{\ba}{\begin{eqnarray}}
\newcommand{\ea}{\end{eqnarray}}
\usepackage{epsf,epsfig,graphics}
\usepackage{verbatim,color,ulem}
\begin{document}
\title{Strong gravitational lensing by Kerr and Kerr-Newman black holes}
\author{Tien Hsieh}
%\email{410414203@gms.ndhu.edu.tw}
\author{Da-Shin Lee}
\email{dslee@gms.ndhu.edu.tw}
\author{Chi-Yong Lin}
\email{lcyong@gms.ndhu.edu.tw}
\affiliation{
Department of Physics, National Dong Hwa University, Hualien 97401, Taiwan, Republic of China}
\date{\today}

\begin{abstract}
We study  the strong gravitational lensing due to  the Kerr black  holes with angular momentum $a$ and the Kerr-Newman black holes with additional charge $Q$.
% respectively.
%
We first derive the analytical expressions of the deflection angles of light rays that particularly diverge as they travel near the photon sphere. In this strong deflection limit, the light rays can circle around the black hole multiple times before reaching the observer, giving relativistic images.
The obtained analytical expressions are then applied to compute the angular positions of relativistic images due to the supermassive  galactic black holes.
In this work, we focus on the outermost image with reference to the optical axis.
%
%\textcolor{blue}
{We find that its angular separation from the one closest to the optical axis increases with the increase of angular momentum $a$ } of the black holes for light rays in direct orbits.
Additionally, the effects of the charge $Q$ of black holes also increase the angular separation of the outermost image from the others  for both direct and retrograde orbits.
The potentially increasing observability of the relativistic images from the effects of angular momentum and charge of the black holes will be discussed.
\end{abstract}


\maketitle

\section{Introduction}
Gravitational lensing is one of the powerful tools to test general relativity (GR) \cite{MIS,HAR}.
Weak lensing  has been fully studied in the formalism of weak field approximations, which can be used to successfully explain various lensing phenomena in a broad array of astrophysical contexts \cite{SEF1992}.
Nevertheless, in recent years, there have been significant theoretical studies looking into lensing  phenomena from strong field perspectives \cite{Vir,Fri,Bozza1,Bozza2,Bozza_2003,Bozza3,Eiroa,Iyer1,Tsuka1,Tsuka2,Vir3,Sha}.
Through the gravitational lensing in the vicinity of the compact massive objects such as a black hole would provide another  avenue to test GR.
So far, observational evidence has shown that almost every large galaxy has a supermassive black hole at the galaxy's center \cite{Ric}.
The Milky Way has a supermassive black hole in its Galactic Center with the location of Sagittarius A* \cite{Ghez,Sch}.
Together with the first image of  the black hole  captured by the Event Horizon Telescope \cite{EHT1,EHT2,EHT3}, gravitational lensing will also become an important probe to study the isolated dim black hole.

\begin{figure}[htp]
\begin{center}
\includegraphics[width=15cm]{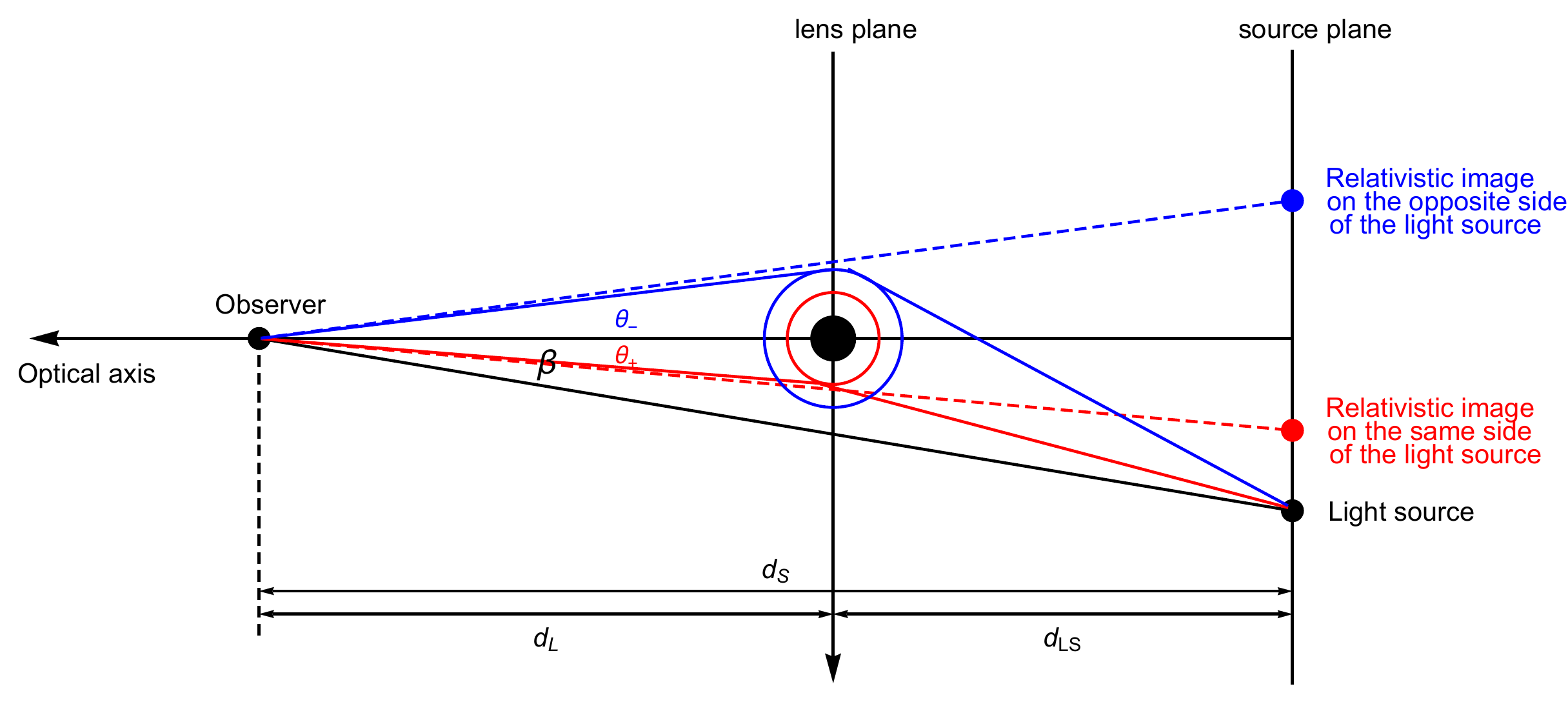}
\caption{Gravitational lens about relativistic images.
         Considering the Kerr or the Kerr-Newman black hole with angular momentum of the clockwise rotation, the light rays are emitted from the source, and circle around the black hole multiple times in the SDL  along a direct orbit (red line) or a retrograde orbit (blue line). The graph illustrates two sets of the relativistic images. }
        \label{fig:arch_02}
    \end{center}
\end{figure}

Recently, Virbhadra and Ellis  have developed a new gravitational lens equation, which allows us to study large deflection of light rays, resulting in the strong gravitational  lensing \cite{Vir}.
This lens equation is then applied to analyze the lensing by a Schwarzschild black hole in the center of the galaxy using numerical methods.
Later, Frittelli {\it et al.} propose the definition of an exact lens equation without reference to the background spacetime, and construct the exact lens equation explicitly in the Schwarzschild spacetime \cite{Fri}.
Strong field lensing in the general spherically symmetric and static spacetime is first studied analytically by Bozza in \cite{Bozza1,Bozza2,Bozza3} and later by Tsukamoto in \cite{Tsuka1,Tsuka2}.
These works show that the deflection angle $\hat{\alpha}(b)$ of light rays for a given impact parameter $b$, which  in the strong deflection limit (SDL) as $b\to b_{c}$, can be approximated in the form
\begin{equation} \label{hatalpha_as}
\hat{\alpha}(b)\approx-\bar{a}\log{\left(\frac{b}{b_{c}}-1\right)}+\bar{b}+O( (b-b_{c}) \log(b-b_{c}) )
\end{equation}
with two parameters $\bar a$ and $\bar b$ as a function of the black hole's parameters.
Then, in \cite{Bozza_2003},  the Kerr black hole of the nonspherically symmetric black holes is considered, exploring $\bar a$ and $\bar b$ numerically.
In this paper, we extend the works of \cite{Bozza1} and \cite{Tsuka1,Tsuka2} and find the analytic form of $\bar a$ and $\bar b$  for nonspherically symmetric  Kerr and Kerr-Newman black holes, respectively, using the analytical closed-form expressions of the deflection angles in \cite{Iyer2} and \cite{Hsiao}.
Although one might not expect that astrophysical black holes have large residue electric charge, some accretion scenarios are proposed to investigate the possibility of the spinning charged back holes \cite{Wilson_1975,Dam}.
It is then still of great interest to extend the studies to the Kerr-Newman black holes \cite{Liu,Jiang_2018, Kraniotis_2014}.
The analytical expressions can be  applied to examine the lensing effects due to the supermassive  galactic black holes  as illustrated in Fig.(\ref{fig:arch_02}).  The light rays are emitted from the source, and circle around the black hole multiple times in the SDL  along a direct orbit (red line) or  a retrograde orbit (blue line), giving two sets of the relativistic images.
Following the approach of \cite{Bozza2} enables us to  study the observational consequences.

The layout of the paper is as follows. In Sec.II, we first review the closed-form expression of the deflection angle due to the Kerr and/or the Kerr-Newman black holes.
% in terms of elliptic integrals.
%
In particular, we discuss the results of the radius of the innermost circular motion of light rays as well as the associated critical impact parameters as a function of the black hole's parameters.  These will  serve as the important inputs to find the values of the coefficients $\bar a$ and $\bar b$ in the SDL deflection angle.
Then we derive the analytic form of $\bar a$ and $\bar b$ in the cases of Kerr and Kerr-Newman black holes, respectively, and check the consistency with the known results from taking the proper limits of the black holes's parameters.
%
%\textcolor{blue}
{In Sec. III, the analytical expressions on the equatorial gravitational lensing   are then applied to compute the angular positions  of relativistic images due to the supermassive  galactic black holes.
When the light rays travel on the quasiequatorial plane, the obtained results for $\theta=\frac{\pi}{2}$ can also be used  to estimate the magnification of relativistic images,
as the light sources are near one of the caustic points with the additional inputs from the  dynamics of the light rays in the  angle $\theta$.}
The potentially increasing observability of the relativistic images from the effects of angular momentum and charge of the black holes will be summarized in the closing section.

\section{Deflection angle due to black holes in the strong deflection limit}

We consider nonspherically symmetric spacetimes of the Kerr  and Kerr-Newman metrics respectively to obtain the deflection angle $\hat{\alpha}(b)$ of light rays for a given impact parameter $b$.
In the SDL, as $b\to b_{c}$, $\hat{\alpha}(b)$ can be approximated in the form of (\ref{hatalpha_as}).
In what follows, we will consider the above two types of the black holes separately.

\subsection{Kerr black holes}

The line element of the Kerr black hole in which spacetime outside a black hole with the gravitational mass $M$ and  angular momentum per unit mass $a=J/M$ is described by
\ba
 {ds}^2 &=& g_{\mu\nu} dx^\mu dx^\nu  \nonumber \\
&=& -\frac{ \left(\Delta -a^2 \sin^2\theta \right)}{\Sigma } {dt}^2   - \frac{ a   \sin ^2\theta \left(2 M
   r\right)}{\Sigma } ({dt}{d\phi+ d\phi dt)}  \nonumber\\
&&
%\quad\quad
+\frac{\Sigma}{\Delta} dr^2 +\Sigma{\, d\theta}^2 +\frac{ \sin ^2 \theta}{\Sigma} \left(( r^2+a^2)^2 -a^2 \Delta  \sin
   ^2\theta \right) {d\phi}^2 \,  \label{KN_metric}
   \ea
with
\be
\Sigma=r^2+a^2\cos^2\theta \, , \quad  \Delta=r^2+a^2-2 M r\,.
\ee
The outer (inner)  event horizon $r_+$ ($r_-$) can be found by solving $\Delta(r)=0$, and is given by
\be \label{rpm_k}
r_{\pm}=M\pm\sqrt{M^2-a^2}\,
\ee
with the condition $M^2 > a^2$. Notice that we just adopt the notation of $r_-$ where the light rays traveling outside the horizon are considered.

The Lagrangian  of a particle is then
\begin{align}
\mathcal{L}=\frac{1}{2}g_{\mu\nu}u^\mu u^\nu \,
\end{align}
with the 4-velocity  $u^\mu=dx^\mu/d\lambda$ defined in terms of an affine parameter $\lambda$.
The metric of the Kerr black hole, which is independent on $t$ and $\phi$,  gives the associated Killing vectors
$\xi_{(t)}^\mu$ and $\xi_{(\phi)}^\mu$
%yielding,
%
\begin{align}
\xi_{(t)}^\mu=\delta_t^\mu \; , \quad \xi_{\phi}^\mu=\delta_\phi^\mu \,.
\end{align}
Then, together with 4-velocity of light rays, the  conserved quantities along a geodesic, can be constructed by the above Killing vectors
$ \varepsilon  \equiv -\xi_{(t)}^\mu u_\mu $ and $\ell  \equiv \xi_{\phi}^\mu u_\mu $,
where $\varepsilon$ and $\ell$ are the light ray's energy and azimuthal angular momentum at spatial infinity.
Light rays traveling along null world lines obey the condition $ u^\mu u_\mu=0$.
To indicate whether the light rays are traversing along the direction of frame dragging or opposite to it, we define the following impact parameter :
 \begin{align}
 b_s=s\left|\frac{\ell}{\varepsilon}\right|\equiv s\,b \; ,
 \end{align}
where $s=\text{Sign$(\ell/\varepsilon)$}$ and $b$ is the positive magnitude.
The parameter $s=+1$ for $b_s>0$ will be referred to as direct orbits, and those with $s=-1$ for $b_s<0$ as retrograde orbits (see Fig.(\ref{fig:arch_02}) for the sign convention).
Here we restrict the light rays traveling on the equatorial plane of the black hole by choosing $\theta={\pi}/{2}$, and $\dot{\theta}=0$.
The equation of motion along the radial direction can be cast in the form  \cite{Hsiao}
 \begin{align} \label{r_eq}
 \frac{1}{b^2} & =\frac{\dot{r}^2}{\ell^2}+W_\text{eff}(r)\; ,
 \end{align}
from which we define the function $W_\text{eff}$ as
 \begin{align}
 W_\text{eff}(r)=\frac{1}{r^2}\left[1-\frac{a^2}{b^2}-\frac{2M }{r}\left(1-\frac{a}{b_s}\right)^2\right] \,
 . \label{eq:Weff_k}
 \end{align}

The above equation is analogous to that of particle motion in the effective potential $ W_\text{eff}(r)$
with the kinetic energy ${\dot r}^2/\ell^2$ and constant total energy $1/b^2$.
Let us consider that a light ray  starts in the asymptotic region to approach the black hole, and then turns back to the asymptotic region to reach the observer.
Such light rays have a turning point, the  closest approach distance to a black hole $r_0$, which  crucially depends on the impact parameter $b$, determined by
 \begin{align}
 \left.\frac{\dot{r}^2}{\ell^2}\right|_{r=r_0}=\frac{1}{b^2}-W_\text{eff}(r_0)=0 \,. \label{eq:rdot_Weff}
 \end{align}
From~\eqref{eq:rdot_Weff},  also shown in \cite{Hsiao,Iyer2}, one can find the impact parameter $b$ for a given $r_0$, which becomes the important input for the analytical expressions of the deflection angle in the SDL, as
\begin{equation} \label{br0_k}
b(r_0)=\frac{2sMa -r_0 \sqrt{a^2-2r_0 M+r_0^2}}{2M-r_0}\,.
\end{equation}
The behavior of the light ray trajectories depends on whether $1/b^2$ is greater or less than the maximum height of $W_\text{eff}(r)$.
The innermost trajectories of light rays  have a direct consequence on  the apparent shape of the black hole.
The smallest radius ${r_{sc}}$, when the turning point $r_0$ is located at the maximum of $W_\text{eff}(r)$, with the critical impact parameter ${b_{sc}}$, obeys
 \begin{align}\label{rsc}
 \left.\frac{d\,W_\text{eff}(r)}{dr}\right|_{r=r_{sc} } & =0  \, .
 \end{align}
Then the radius of the circular motion forming the photon sphere is given by (See \cite{Hsiao,Iyer2}).
\begin{align} \label{rsc_k}
r_{sc}=  2 M\bigg\{ 1+ \cos\bigg[\frac{2}{3} \cos^{-1} \bigg( \frac{-sa}{M} \bigg) \bigg]\bigg\}
\;
\end{align}
with the corresponding impact parameter
\begin{align} \label{bsc_k}
 b_{sc} & =-a+ s6 M \cos\bigg[\frac{1}{3} \cos^{-1} \bigg( \frac{-sa}{M} \bigg) \bigg] \, .
 \end{align}

In the case of a Kerr black hole, the nonzero spin of the black hole is found to give more repulsive  effects  to the light rays in the direct  orbits than those in the retrograde orbits due to  the $1/r^3$ term in the effective potential.
%, as seen from its contributions to the
%
The repulsive  effects in turn affect light rays  in the direct orbits in a way  to prevent them from collapsing into the event horizon.
%and thus as will be seen later,
 As a result, this shifts the innermost   circular trajectories  of the light rays  toward the black hole with the smaller critical impact parameter $b_{+c}$ than  $b_{-c}$ in the retrograde orbits as shown in Fig.(\ref{bsc_a}).
As such, when $a$ increases, the impact parameter $b_{+c}$ decreases whereas $\vert b_{-c}\vert$ increases instead \cite{Iyer2,Hsiao}.
It will be shown in the next section that the value of $ b_{sc} $ is a key quantity to determine the features of the angular position of the induced images of the distant light sources due to the strong  gravitational lensing effects.
Also, the presence of black hole's spin  is to give the smaller  deflection angle in the direct orbits as  compared with the retrograde orbits with the  same impact parameter $b$ \cite{Hsiao,Iyer2}.

We proceed by introducing the variable
\begin{equation}\label{z}
z \equiv 1-\frac{r_0}{r} \, .
\end{equation}
The geodesic equations for $r$ and $\phi$ found in \cite{Hsiao} can be rewritten in terms of $z$ as \cite{Tsuka1}
\begin{equation}\label{dz/dphi}
\frac{dz}{d\phi}=\frac{1}{r_0}\frac{1-\frac{2{M}}{r_0}(1-z)+\frac{a^2}{r_0^2}(1-z)^2}
                  {1-\frac{2{M}}{r_0}(1-z)(1-\frac{a}{b_s})}\sqrt{B(z,r_0)}\;,
\end{equation}
where the function $B(z,r_0)$ has the trinomial form in $z$
\begin{equation}
B(z,r_0)=c_1(r_0)z+c_2(r_0)z^2+c_3(r_0)z^3
\end{equation}
with the coefficients
\begin{equation}
\begin{split} \label{cs_k}
c_1(r_0)=&-6Mr_0\left(1-\frac{a}{b_s}\right)^2+2r_0^2\left(1-\frac{a^2}{b^2}\right)\,,\\
c_2(r_0)=&6Mr_0\left(1-\frac{a}{b_s}\right)^2-r_0^2\left(1-\frac{a^2}{b^2}\right)\, ,\\
c_3(r_0)=&-2Mr_0\left(1-\frac{a}{b_s}\right)^2 \, .
\end{split}
\end{equation}
Next we rewrite
\begin{equation}
\frac{1-\frac{2{M}}{r_0}(1-\frac{a}{b_s})+\frac{2{M}}{r_0}(1-\frac{a}{b_s})z}
{ 1-\frac{2M}{r_0}+\frac{a^2}{r_0^2}+(\frac{2{M}}{r_0}-\frac{2a^2}{r_0^2})z+\frac{a^2}{r_0^2}z^2}
=\frac{r_0^2}{a^2}\left(\frac{C_-}{z-z_-}+\frac{C_+}{z-z_+}\right)\;,
\end{equation}
where the roots $z_-$, $z_+$, and the coefficients $C_-$, $C_+$ are
\begin{equation}
\begin{split} \label{zpm_k}
z_-=&1-\frac{r_0r_-}{a^2}\, ,\\
z_+=&1-\frac{r_0r_+}{a^2} \, ,
\end{split}
\end{equation}
\begin{equation}\label{Cpm_k}
\begin{split}
C_-=&\frac{a^2-2Mr_-(1-\frac{a}{b_s})}{2r_0 \sqrt{M^2-a^2}}\, ,\\
C_+=&\frac{-a^2+2Mr_+(1-\frac{a}{b_s})}{2r_0 \sqrt{M^2-a^2}}
\end{split}
\end{equation}
with  $r_+$ ($r_-$) being the outer (inner) horizon of a Kerr black hole defined in (\ref{rpm_k}).
Also note that $z_-$, $z_+ \le 0$, for  all spin $a$.
%where $-1< a <1$.
Then the deflection angle can be calculated as a function of the closest approach distance $r_0$ from (\ref{dz/dphi}) giving
\begin{equation}\label{alpha_I}
\hat\alpha(r_0) = I (r_0) -\pi \, , \quad I(r_0)= \int_0^1 f(z,r_0) dz \, ,
\end{equation}
where the integrand becomes
\begin{equation}
 f(z,r_0)= \frac{r_0^2}{a^2}\left(\frac{C_-}{z-z_-}+\frac{C_+}{z-z_+}\right)\frac{2{r_0}}{\sqrt{c_1(r_0)z+c_2(r_0)z^2+c_3(r_0)z^3}} \,.
\end{equation}
In the SDL  of our interest, when the closest approach distance reaches  its critical limit, namely  $r_0 \to r_{sc}$, and $c_1(r_0) \to 0$ in (\ref{cs_k}) obtained from (\ref{rsc}), the integrand  $f(z,r_0) \rightarrow \frac{1}{z}$ for small $z$ leads to the logarithmic divergence as  $r_0 \to r_{sc}$.
%resulting in   (\ref{hatalpha_as}).
Let us now define a new function $f_D(z,r_0)$
\begin{equation}
f_D(z,r_0)=\frac{r_0^2}{a^2}\left(\frac{C_-}{z-z_-}+\frac{C_+}{z-z_+}\right)\frac{2{r_0}}{\sqrt{c_1(r_0)z+c_2(r_0)z^2}}\;,
\end{equation}
that separates the divergent part from  the regular  part given by $f_R(z,r_0)=f(z,r_0)-f_D(z,r_0)$. The integral of $f_R$ is thus finite.

The divergent part comes from an integral of the function $f_D(z,r_0)$, which contributes not only to $\bar a$ for the logarithmic term but also $\bar b$ for the regular part in  (\ref{hatalpha_as}), giving
\begin{equation} \label{I_D_k}
\begin{split}
I_D(r_0)=&\int_0^1 f_D(z,r_0) dz\\
=&\frac{2r_0^3}{a^2}\frac{C_-}{\sqrt{c_1(r_0)z_-+c_2(r_0)z_-^2}}
\log{\left(\frac{\sqrt{c_1(r_0)z_-+c_2(r_0)z_-}+\sqrt{c_1(r_0)+c_2(r_0)z_-}}
{\sqrt{c_1(r_0)z_-+c_2(r_0)z_-}-\sqrt{c_1(r_0)+c_2(r_0)z_-}} \right)} \\
+&\frac{2r_0^3}{a^2}\frac{C_+}{\sqrt{c_1(r_0)z_++c_2(r_0)z_+^2}}
\log{\left(\frac{\sqrt{c_1(r_0)z_++c_2(r_0)z_+}+\sqrt{c_1(r_0)+c_2(r_0)z_+}}
{\sqrt{c_1(r_0)z_++c_2(r_0)z_+}-\sqrt{c_1(r_0)+c_2(r_0)z_+}} \right)} \, .
\end{split}
\end{equation}
In the SDL, the expansions of the coefficient $c_1(r_0)$ (\ref{cs_k}) and the impact parameter $b(r_0)$ in powers of small $r_0-r_{sc}$ read
\begin{equation}
c_1(r_0)=c_{1 sc}' (r_0-r_{sc})+{O}(r_0-r_{sc})^2 \, ,\label{c_SDL}
\end{equation}
\begin{equation} \label{b_SDL}
b(r_0) = b_{sc}+\frac{b_{sc}''}{2!}(r_0-r_{sc})^2+{O}(r_0-r_{sc})^3 \, ,
\end{equation}
where $c_1(r_{sc})\equiv c_{1sc}=0$ and $b(r_{sc})\equiv b_{sc}$ is the critical impact parameter given by (\ref{bsc_k}).
The subscript $sc$ denotes evaluating the function at $r=r_{sc}$. The prime means the derivative with respect to $r_0$.  Notice that using $c_{1sc}=0$ in (\ref{cs_k}), one finds
\begin{equation}\label{c2c3_k}
 c_{3sc}=-\frac{2}{3}c_{2sc}\;.
 \end{equation}
Combining (\ref{c_SDL}) with (\ref{b_SDL}),  we can write  $c_1(r_0)$  in terms of small $b-b_{sc}$ as
\begin{equation} \label{c_b_SDL}
\lim_{r_0\to r_{sc}} c_1(r_0)=\lim_{b\to b_{sc}}c_{1 sc}'\sqrt{\frac{2b_{sc}}{b_{sc}''}}\left(\frac{b}{b_{sc}}-1\right)^{1/2}\;.
\end{equation}
In the SDL, substituting (\ref{c_b_SDL}) into (\ref{I_D_k}), $I_D$ becomes
\begin{equation} \label{ID}
\begin{split}
I_D(b)\simeq &-\left(\frac{r_{sc}^3}{a^2}\frac{C_{- sc}}{\sqrt{c_{2 sc} \, z_{- sc}^2}}+\frac{r_{sc}^3}{a^2}\frac{C_{+ sc}}{\sqrt{c_{2 sc}\, z_{+sc}^2}} \right)\log{\left(\frac{b}{b_{sc}}-1\right)}\\
&+\frac{r_{sc}^3}{a^2}\frac{C_{- sc}}{\sqrt{c_{2 sc}\, z_{-sc}^2}}\log{\left(\frac{16 \, c^2_{2 sc} \, z_{-sc}^2 b_{sc}''}{c_{1sc}'^2 2b_{sc}(z_{- sc}-1)^2}\right)}
+\frac{r_{sc}^3}{a^2}\frac{C_{+sc}}{\sqrt{c_{2 sc}z_{+sc}^2}}\log{\left(\frac{16\, c^2_{2sc} \, z_{+sc}^2 \, b_{sc}''}{c_{1sc}'^2 2b_{sc}(z_{+sc}-1)^2}\right)} \, .
\end{split}
\end{equation}
Finally,  the coefficients $\bar a$ and the contribution from $I_D(b)$ to $\bar b$ denoted by $b_D$ in (\ref{hatalpha_as}) are
\begin{equation}
\begin{split} \label{abar_k}
\bar{a}=&\frac{r_{sc}^3}{\sqrt{c_{2 sc}
}}\left[ \frac{C_{-sc}}{r_{sc}r_{-}-a^2}+\frac{C_{+sc}}{r_{sc}r_{+}-a^2}\right]
\end{split}
\end{equation}
and
\begin{equation}
\begin{split} \label{bD}
b_D=
%&
\bar{a} \log{\left[ \frac{8c_{2sc}^2 b_{sc}''}{c_{1sc}'^2 b_{sc}}\right]}
%\\
%&
+\frac{ 2 r_{sc}^3}{\sqrt{c_{2sc}}}\left[ \frac{C_{-sc}}{r_{sc}r_{-} -a^2}\log{\left(1-\frac{a^2}{r_{sc}r_{-}}\right)}+\frac{C_{+sc} }{r_{sc}r_{+}-a^2}\log{\left(1-\frac{a^2}{r_{sc}r_{+}}\right)} \right] \, ,
\end{split}
\end{equation}
where $z_{\pm}$ are replaced by  $r_{\pm}$ through (\ref{zpm_k}).
The leading order result in the SDL  from the integration of $ f_R(z,r_{sc})$, which contributes the coefficient $\bar b$, is denoted by $b_R$, and is obtained as
\begin{equation}
\begin{split} \label{bR_k}
b_R& =I_R(r_{sc})=\int_0^1 f_R(z,r_{sc}) dz\\
=&\frac{2 r_{0}^3}{a^2}\frac{C_-}{\sqrt{c_2}z_-}\log{\left(\frac{z_-}{z_--1} \frac{\sqrt{c_2+c_3}+\sqrt{c_2}}{\sqrt{c_2+c_3}-\sqrt{c_2}} \frac{c_3}{4c_2} \right)}\\
&+\frac{2 r_{0}^3}{a^2}\frac{C_-}{\sqrt{c_2+c_3z_-}z_-}\log{\left( \frac{\sqrt{c_2+c_3z_-}-\sqrt{c_2+c_3}} {\sqrt{c_2+c_3z_-}+\sqrt{c_2+c_3}} \frac{\sqrt{c_2+c_3z_-}+\sqrt{c_2}}{\sqrt{c_2+c_3z_-}-\sqrt{c_2}}\right)}\\
&+\frac{2 r_{0}^3}{a^2}\frac{C_+}{\sqrt{c_2}z_+}\log{\left(\frac{z_+}{z_+-1} \frac{\sqrt{c_2+c_3}+\sqrt{c_2}}{\sqrt{c_2+c_3}-\sqrt{c_2}} \frac{c_3}{4c_2} \right)}\\
&+\frac{2 r_{0}^3}{a^2}\frac{C_+}{\sqrt{c_2+c_3z_+}z_+}\log{\left( \frac{\sqrt{c_2+c_3z_+}-\sqrt{c_2+c_3}} {\sqrt{c_2+c_3z_+}+\sqrt{c_2+c_3}} \frac{\sqrt{c_2+c_3z_+}+\sqrt{c_2}}{\sqrt{c_2+c_3z_+}-\sqrt{c_2}}\right)}\Big\vert_{r_0=r_{sc}}  \;.
\end{split}
\end{equation}
Thus, the coefficient $\bar b$ can be computed from the sum of  $b_D$ and $b_R$
\begin{equation}
\bar{b}=-\pi+b_D+b_R \,
\end{equation}
with the help of (\ref{bD}) and (\ref{bR_k}). In (\ref{bR_k}) we again use (\ref{c2c3_k}) and (\ref{zpm_k}) to replace $c_{3sc}$ by $c_{2sc}=-\frac{2}{3} c_{3 sc}$ and $z_{\pm}$ by $r_{\pm}$. After some straightforward algebra we find
%
%where
%it is straightforward to find that
\begin{equation}
\begin{split}\label{bbar_k}
\bar{b}=&-\pi+\bar{a} \log{\left( \frac{36}{7+4\sqrt{3}}\frac{8c_{2sc}^2 b_{sc}''}{c_{1sc}'^2 b_{sc}} \right)}\\
&+\frac{r_{sc}^3}{\sqrt{c_{2 sc}}}\frac{2 a C_{-sc}}{ a^2-r_{sc}r_- } \frac{\sqrt{3}}{\sqrt{a^2+2r_{sc}r_-}} \log{\left( \frac{\sqrt{a^2+2r_{sc}r_-}-a}{\sqrt{a^2+2r_{sc}r_-}+a} \frac{\sqrt{a^2+2r_{sc}r_-}+\sqrt{3}a}{\sqrt{a^2+2r_{sc}r_-}-\sqrt{3}a}\right)}\\
&+\frac{r_{sc}^3}{\sqrt{c_{2sc}}}\frac{2 a C_{+sc}}{ a^2-r_{sc}r_+ } \frac{\sqrt{3}}{\sqrt{a^2+2r_{sc}r_+}} \log{\left( \frac{\sqrt{a^2+2r_{sc}r_+}-a}{\sqrt{a^2+2r_{sc}r_+}+a} \frac{\sqrt{a^2+2r_{sc}r_+}+\sqrt{3}a}{\sqrt{a^2+2r_{sc}r_+}-\sqrt{3}a}\right)} \, .
\end{split}
\end{equation}
Using the results  of $r_{sc}$ (\ref{rsc_k}), $b_{sc}$ (\ref{bsc_k}) and the expression of $b (r_0)$ (\ref{br0_k}), together with the definitions of $C_\pm$ and $c_2$ in  (\ref{Cpm_k}) and (\ref{cs_k}) respectively, one can compute the coefficients  $\bar a$ and $\bar b$ given by (\ref{abar_k}) and  (\ref{bbar_k})  in the form of (\ref{hatalpha_as}). Notice that with the parameters under investigation $\bar a>0$, but $\bar b<0$.
Our results are shown in Fig.(\ref{ab_k_f}), where  both $\bar a$ and $\vert \bar b \vert$ increase (decrease) in $a$ in  direct (retrograde) orbits, giving the fact that the deflection angle $\hat \alpha$ decreases (increases) with the increase of the black hole's spin for a given impact parameter.
Later in Sec. III we will  compare with the full numerical computations from (\ref{alpha_I}) in the SDL.

The results of $\bar a$ and $\bar b$ due to the  Schwarzschild black hole  in \cite{Bozza2,Tsuka1} can be reproduced  by sending $a \to 0$ where
$ r_+ \rightarrow 2M$, $ r_- \to a^2/2M$,
$ C_{+sc} \to 2M/r_{sc}$,
$ C_{-sc}\to a^3/2b_{sc}Mr_{sc}$, and
 $c_{2sc} \to r_{sc}^2$ using $c_{1sc}=0$ in (\ref{rpm_k}) and (\ref{Cpm_k}).
We can check that $\bar{a}=1$ in (\ref{abar_k}) and $\bar{b}$ in (\ref{bbar_k}) reduces to the expression proportional to $\bar a$ given by
\begin{equation}
\begin{split}
\bar{b}=&-\pi+\bar{a} \log{\left( 36(7-4\sqrt{3})\frac{8c_{2 sc}^2 b_{sc}''}{c_{1sc}'^2 b_{sc}} \right)}\\
=&-\pi+ \log{\left( 216(7-4\sqrt{3}) \right)}\, .
\end{split}
\end{equation}
In the second equality above we have  further used substitutions $ b_{sc} \to 3\sqrt{3}M$, $b_{sc}''\to\sqrt{3}/M$,  $c_{1sc}' \to 6 M$, and $  c_{2sc} \to 9M^2$ obtained from $r_{sc}=3M$ in the  Schwarzschild black hole.
In Fig.(\ref{alpha_com}), we compare the approximate results of the deflection angle in the SDL with the exact ones in \cite{Iyer2} and \cite{Hsiao}, and find that they  are in good agreement when $b\to b_{sc}$.

The analytical expressions of the coefficient $\bar a$ and $\bar b$ in the form of the SDL deflection angle  due to the Kerr black hole are successfully achieved. They are an extension of the works in \cite{Bozza3} and \cite{Tsuka1} where the spherically symmetric black holes are considered. This is one  of the main results in this work.

\subsection{Kerr-Newman black holes}

We now consider another example with the nonspherically symmetric  metric of a charged spinning black hole. With an addition of charge $Q$ comparing with the Kerr case, the line element associated with the Kerr-Newman metric  is
\ba
 {ds}^2 &=& g_{\mu\nu} dx^\mu dx^\nu  \nonumber \\
&= & -\frac{ \left(\Delta -a^2 \sin^2\theta \right)}{\Sigma } {dt}^2   + \frac{ a   \sin ^2\theta \left(Q^2-2 M
   r\right)}{\Sigma } ({dt}{d\phi+ d\phi dt)}  \nonumber\\
&&+  \frac{\Sigma}{\Delta} dr^2 +\Sigma{\, d\theta}^2 +\frac{ \sin ^2 \theta}{\Sigma} \left(( r^2+a^2)^2 -a^2 \Delta  \sin
   ^2\theta \right) {d\phi}^2 \, , \label{KN_metric}
   \ea
where
\be
\Sigma=r^2+a^2\cos^2\theta \, , \quad  \Delta=r^2+a^2+Q^2-2 M r\,.
\ee
The outer (inner) event horizon $r_+$ ($r_-$) is
\be \label{rpm_kn}
r_{\pm}=M\pm\sqrt{M^2-(Q^2+a^2)}\,
\ee
with  $M^2 > Q^2+ a^2$.

\begin{figure}[htp]
\centering
\includegraphics[width=0.88\columnwidth=0.88]{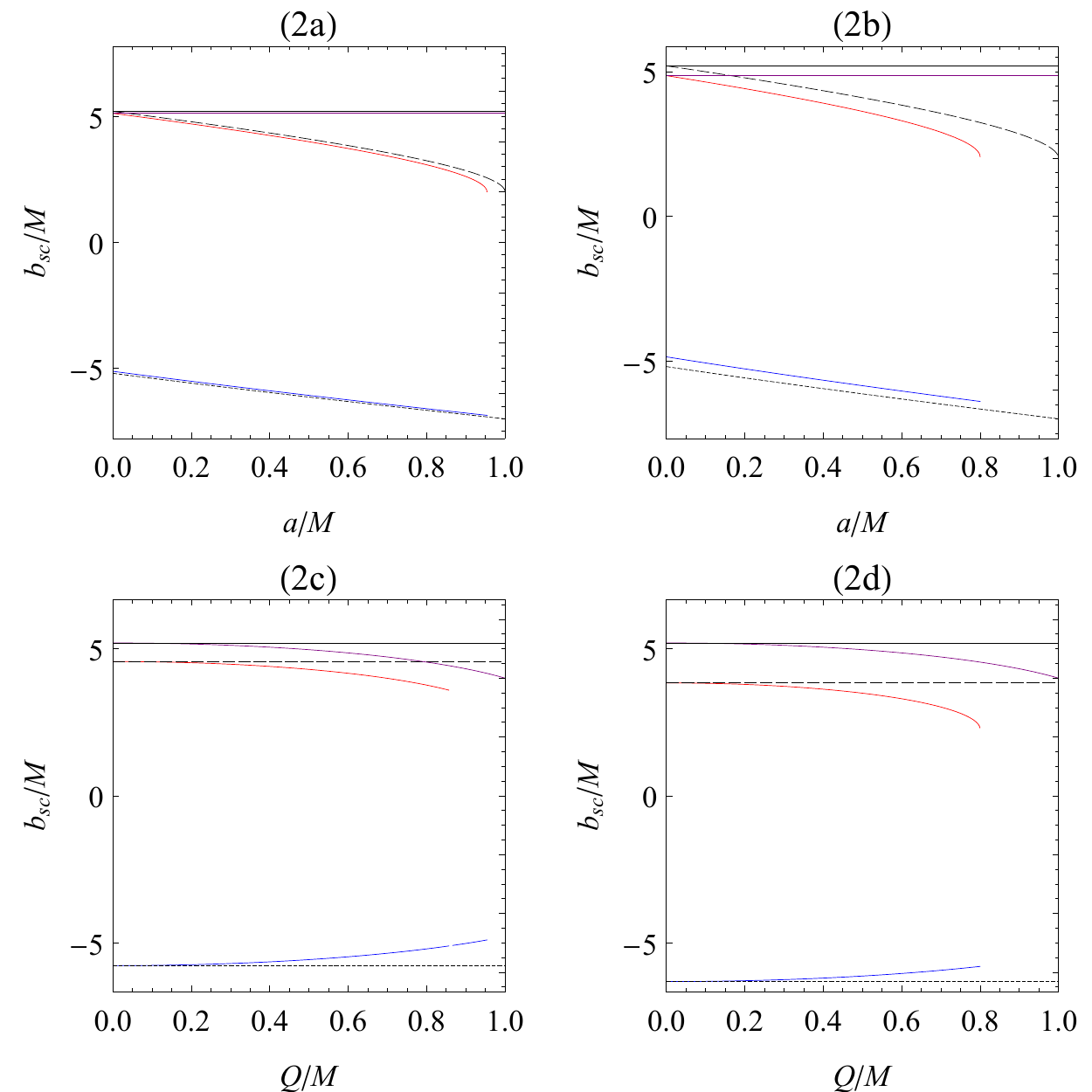}
\caption{
 %\textcolor{blue}
The critical impact parameter $b_{sc}/M$ as a function of the spin parameter $a/M$ for (a) $Q/M=0.3$, (b) $Q/M=0.6$.
Also, the critical impact parameter $b_{sc}/M$ as a function of charge $Q/M$ for (c) ${a/M}=0.3$, (d) ${a/M}=0.6$.
The plots show the Schwarzschild, Reissner-Nordstr\"om, Kerr and Kerr-Newman black holes for comparison. {The plot convention used henceforth: Kerr-Newman direct (solid red line), Kerr-Newman retrograde (solid blue line), Kerr direct (black dashed line with $Q=0$), Kerr retrograde (black dotted line, with $Q=0$), Reissner-Nordstr\"om (solid purple line, with $a=0$), and Schwarzschild (solid black line, with $Q=0, a=0$).
}
  }
  \label{bsc_a}
\end{figure}

\begin{figure}[h]
\centering
\includegraphics[width=0.88\columnwidth=0.88]{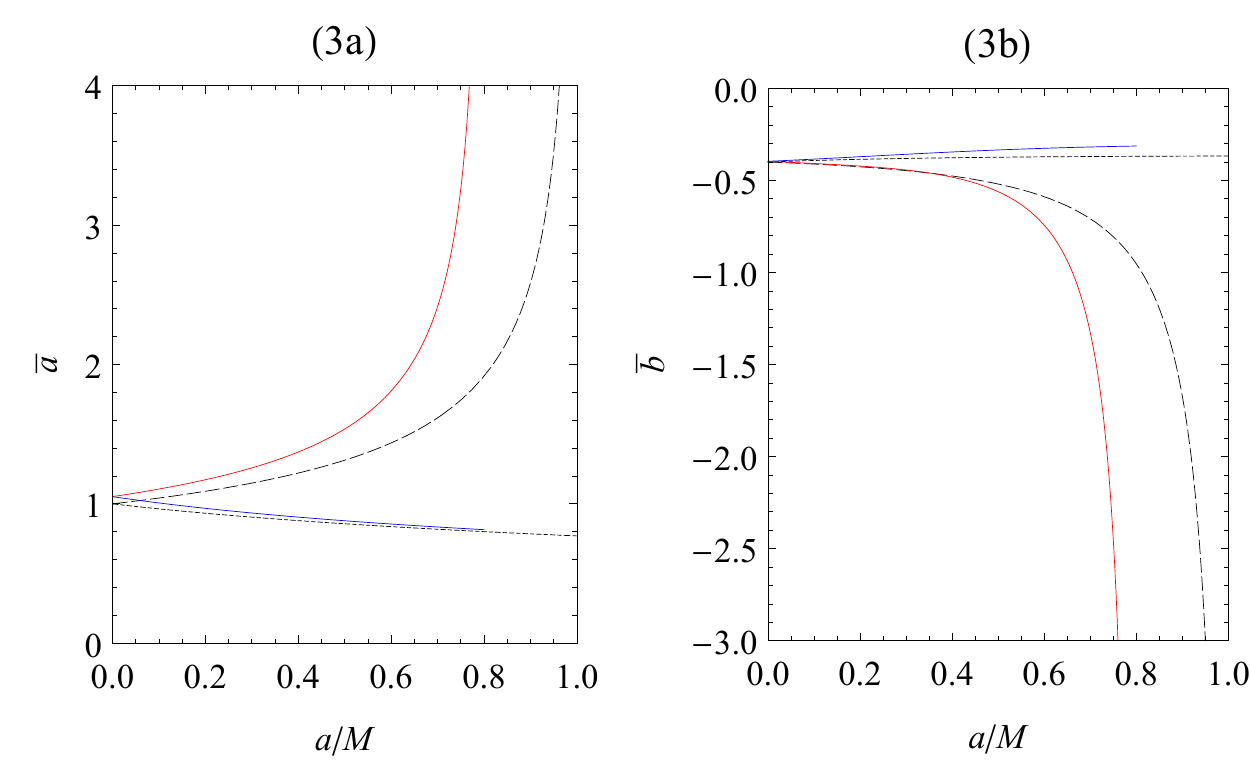}
 \caption{
%\textcolor{blue}
{
%\textcolor{blue}
{The coefficients  $\bar a$ and $\bar b$  as a function of the spin parameter $a/M$ for the Kerr black hole with $Q/M=0$ and the Kerr-Newman black hole with $Q/M=0.6$: (a) the coefficient $\bar a$, (b) the coefficient $\bar b$. The display of the plot follows the convention in Fig.(\ref{bsc_a}).}}}
\label{ab_k_f}
\end{figure}

The light rays traveling on the equatorial plane of the black hole have been studied analytically in our previous work in \cite{Hsiao}, in which  the function $W_\text{eff}$ from the equation of motion along the radial direction in (\ref{r_eq}) can be regarded as an effective potential given by
 \begin{align}
W_\text{eff}(r)=\frac{1}{r^2}
\left[1-\frac{a^2}{b^2}+\left(-\frac{2M}{r}+\frac{Q^2}{r^2}\right)\left(1-\frac{a}{b_s}\right)^2\right] \, . \label{eq:Weff_kn}
\end{align}
For the  Kerr-Newman black hole, the nonzero charge of the black hole is found to give repulsive effects  to light rays as seen from its contributions to the function $W_\text{eff}$ of  the $1/r^4$ term, which shifts  the innermost circular trajectories  of the light rays toward the black holes with the smaller critical impact parameter $b_{sc}$ for both direct and retrograde orbits, as illustrated in Fig.(\ref{bsc_a}).
Also, the presence of black hole's charge is to decrease the deflection angle due to the additional repulsive effects on the light rays, as  compared with the Kerr case with the same impact parameter $b$ \cite{Hsiao}. As we will discuss in the next section, the angular positions of the relativistic  images of the distant light sources due to the gravitational lensing of the black holes critically depends on the critical impact parameter $b_{sc}$.

The impact parameter $b$ as a function of the radius of the circular motion $r_0$ is obtained as
\begin{eqnarray}\label{br0_kn}
b(r_0)=\frac{s(2aM-a\frac{Q^2}{r_0})- r_0 \sqrt{(a\frac{Q^2}{r_0^2}-a\frac{2M}{r_0})^2
+(1-\frac{2M}{r_0}+\frac{Q^2}{r_0^2})[a^2(1+\frac{2M}{r_0}-\frac{Q^2}{r_0^2})+r_0^2]}}
{ 2M-r_0-\frac{Q^2}{r_0}} \, .
\end{eqnarray}
The solution of $r_{sc}$ of the radius of the innermost circular motion has been found in \cite{Hsiao} as
\begin{align}\label{rsc_kn}
 r_{sc} & =\frac{3M}{2}+\frac{1}{2\sqrt{3}}\sqrt{9M^2-8Q^2+U_{c}+\frac{P_{c}}{U_{c}}} \no\\
 &\quad -\frac{s}{2}\sqrt{6M^2-\frac{16Q^2}{3}-\frac{1}{3}\left(U_c+\frac{P_{c}}{U_{c}}\right)
 +\frac{8\sqrt{3}Ma^2}{\sqrt{9M^2-8Q^2+U_c+\frac{P_c}{U_c}}}} \;\; ,
 \end{align}
where
 \begin{align}
 P_{c} & =(9M^2-8Q^2)^2-24a^2(3M^2-2Q^2) \, , \no\\
 U_{c} & =\bigg\{(9M^2-8Q^2)^3-36a^2(9M^2-8Q^2)(3M^2-2Q^2)+216M^2a^4 \no\\
 &\quad\quad +24\sqrt{3}a^2\sqrt{(M^2-a^2-Q^2)\left[Q^2(9M^2-8Q^2)^2-27M^4a^2\right]}\bigg\}^\frac{1}{3} \, .
 \end{align}
The analytical expression of the critical value of the impact parameter ${b_{sc} }$ can be written as a function of black hole's parameters \cite{Hsiao},
% giving the innermost circular motion is obtained to be
\begin{align} \label{bsc_kn}
 b_{sc} & =-a+\frac{M^2a}{2(M^2-Q^2)}+\frac{s}{2\sqrt{3}(M^2-Q^2)}\Bigg[\sqrt{V+(M^2-Q^2)\left(U+\frac{P}{U}\right)} \no\\
 &
 +\sqrt{2V-(M^2-Q^2)\left(U+\frac{P}{U}\right)-\frac{s6\sqrt{3}M^2a\left[(M^2-Q^2)(9M^2-8Q^2)^2-M^4a^2\right]}
 {\sqrt{V+(M^2-Q^2)\left(U+\frac{P}{U}\right)}}}\Bigg]\;,
 \end{align}
where
 \begin{align}
 P & =(3M^2-4Q^2)\left[9(3M^2-4Q^2)^3+8Q^2(9M^2-8Q^2)^2-216M^4a^2\right] \, , \no\\
 U & =\bigg\{-\left[3(3M^2-2Q^2)^2-4Q^4\right]\left[9M^2(9M^2-8Q^2)^3-8\left[3(3M^2-2Q^2)^2-4Q^4\right]^2\right] \, \no\\
 &\qquad +108M^4a^2\left[9(3M^2-4Q^2)^3+4Q^2(9M^2-8Q^2)^2-54M^4a^2\right]\,  \no\\
 &\qquad +24\sqrt{3}M^2\sqrt{(M^2-a^2-Q^2)\left[Q^2(9M^2-8Q^2)^2-27M^4a^2\right]^3}\bigg\}^\frac{1}{3} \, ,\no\\
 V & =3M^4a^2+(M^2-Q^2)\left[6(3M^2-2Q^2)^2-8Q^4\right] \,.
 \end{align}
These will serve as the important inputs for the analytical expressions of the coefficients $\bar a$ and $\bar b$ in (\ref{hatalpha_as}).

The counterpart of  (\ref{dz/dphi}) for the Kerr-Newman case as a function of $z$  in (\ref{z}) can be easily derived giving
\begin{equation}
\frac{dz}{d\phi}=\frac{1}{r_0} \frac{1-\frac{2{M}}{r_0}(1-z)+\frac{a^2+Q^2}{r_0^2}(1-z)^2}
{1-\frac{2{M}}{r_0}(1-\frac{a}{b_s})(1-z)+\frac{Q^2}{r_0^2}(1-\frac{a}{b_s})(1-z)^2} \sqrt{B(z,r_0)}\;,
\end{equation}
where
\begin{equation}
B(z,r_0)=c_1(r_0)z+c_2(r_0)z^2+c_3(r_0)z^3+c_4(r_0)z^4 \, .
\end{equation}
The function $B(z,r_0)$ is then  the quartic polynomial in $z$ with the coefficients
\begin{equation}\label{cs_kn}
\begin{split}
c_1(r_0)=&4Q^2\left(1-\frac{a}{b_s}\right)^2-6Mr_0\left(1-\frac{a}{b_s}\right)^2+2r_0^2\left(1-\frac{a^2}{b^2}\right)\,,\\
c_2(r_0)=&-6Q^2\left(1-\frac{a}{b_s}\right)^2+6Mr_0\left(1-\frac{a}{b_s}\right)^2-r_0^2\left(1-\frac{a^2}{b^2}\right)\, ,\\
c_3(r_0)=&4Q^2\left(1-\frac{a}{b_s}\right)^2-2Mr_0\left(1-\frac{a}{b_s}\right)^2\,,\\
c_4(r_0)=&-Q^2\left(1-\frac{a}{b_s}\right)^2 \, .
\end{split}
\end{equation}
All coefficients have the additional contributions from the charge $Q$. In particular, the presence of the $z^4$ term with the coefficient $c_4(r_0)$ in $B$, which vanishes in  the Kerr case, makes the calculations of $\bar a$ and $\bar b$ more involved.
The integrant function $f(z,r_0)$ in (\ref{alpha_I}) now takes the form
\begin{equation}
%\begin{split}
f(z,r_0)=\frac{r_0^2}{a^2+Q^2}\left(\frac{C_-}{z-z_-}+\frac{C_Q z+C_+}{z-z_+}\right)
%\\
%&\times
\frac{2{r_0}}{\sqrt{c_1(r_0)z+c_2(r_0)z^2+c_3(r_0)z^3+c_4(r_0)z^4}} \,.
%\end{split}
\end{equation}
The corresponding coefficients $C_-$, $C_Q$, and $C_+$ in the Kerr-Newman case are
\begin{equation} \label{CpmQ_kn}
\begin{split}
C_-=&\frac{a^2+Q^2-2Mr_-(1-\frac{a}{b_s})+\frac{Q^2r_-^2}{a^2+Q^2}(1-\frac{a}{b_s})}{2r_0\sqrt{M^2-a^2-Q^2}}\, ,\\
C_Q=&\frac{Q^2}{r_0^2}\left(1-\frac{a}{b_s}\right)\, ,\\
C_+=&\frac{a^2+Q^2-2Mr_-(1-\frac{a}{b_s})+\frac{Q^2}{r_{0}}(r_+-r_-)(1-\frac{a}{b_s})+Q^2(1-\frac{a}{b_s})}
{-2r_{0}\sqrt{M^2-a^2-Q^2}} \, ,
\end{split}
\end{equation}
where  $z_+$, $z_-$ then become
\begin{equation}
\begin{split}
z_-=&1-\frac{r_0r_-}{a^2+Q^2}\;,\\
z_+=&1-\frac{r_0r_+}{a^2+Q^2}\;,
\end{split}
\end{equation}
defined in terms of  the outer(inner) black hole horizon $r_+$ ($r_-$). Again, $z_\pm \le 0$ for all $a$ and $Q$ with the nonzero $r_+$. Note that, for charge $Q \to 0$, $C_Q$ vanishes.

\begin{figure}[h]
\centering
\includegraphics[width=0.88\columnwidth=0.88]{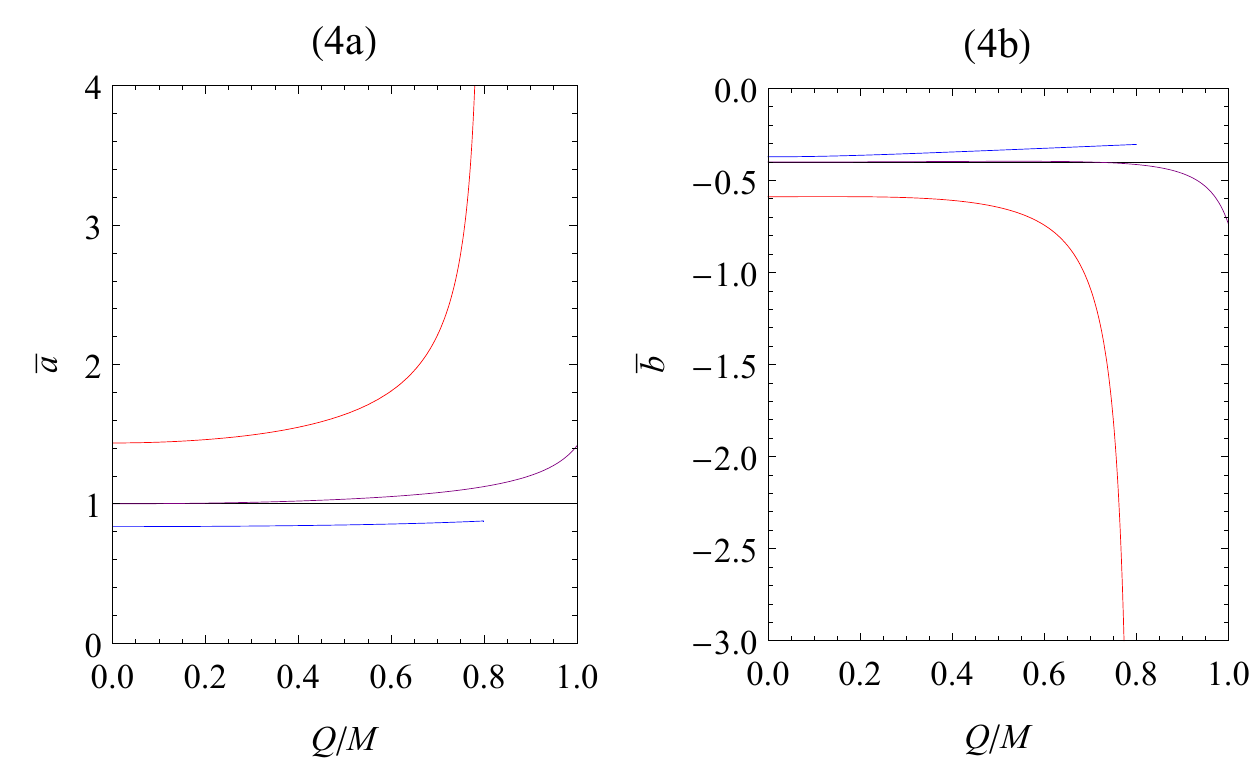}
 \caption{
%\textcolor{blue}
%\textcolor{blue}
{The coefficients of $\bar a$ and $\bar b$  as a function of charge $Q/M$ for the Schwarzschild black hole with  ${a/M}=0$, ${Q/M}=0$,
the the Reissner-Nordstr\"om blck hole with
 ${a/M}=0$  and the Kerr-Newman black hole with ${a/M}=0.6$: (a) the coefficient $\bar a$, (b) the coefficient $\bar b$. The display of the plot follows the convention in Fig.(\ref{bsc_a}). }}
\label{ab_kn_f}
\end{figure}

Analogous to the previous subsection of the Kerr case, we define the function $f_D(z,r_0)$ as
\begin{equation}
f_D(z,r_0)=\frac{r_0^2}{a^2+Q^2}\left(\frac{C_-}{z-z_-}+\frac{C_Q z+C_+}{z-z_+}\right)\frac{2{r_0}}{\sqrt{c_1(r_0)z+c_2(r_0)z^2}}\, .
\end{equation}
As $z\to 0$, $f_D(z,r_0) \to 1/z$. Its integration over $z$ gives the divergent part of $I_D(r_0)$ when $b \to b_c$.
Here we find
\begin{equation}
\begin{split}
I_D(r_0)=&\int_0^1 f_D(z,r_0) dz\\
=&\frac{2r_0^3}{a^2+Q^2}\frac{C_-}{\sqrt{c_1(r_0)z_-+c_2(r_0)z_-^2}}
\log{\left(\frac{\sqrt{c_1(r_0)z_-+c_2(r_0)z_-}+\sqrt{c_1(r_0)+c_2(r_0)z_-}}
{\sqrt{c_1(r_0)z_-+c_2(r_0)z_-}-\sqrt{c_1(r_0)+c_2(r_0)z_-}} \right)}\\
&+\frac{2r_0^3}{a^2+Q^2}\frac{C_++C_Q z_+}{\sqrt{c_1(r_0)z_++c_2(r_0)z_+^2}}
\log{\left(\frac{\sqrt{c_1(r_0)z_++c_2(r_0)z_+}+\sqrt{c_1(r_0)+c_2(r_0)z_+}}
{\sqrt{c_1(r_0)z_++c_2(r_0)z_+}-\sqrt{c_1(r_0)+c_2(r_0)z_+}} \right)}\\
&-\frac{2r_0^3}{a^2+Q^2}\frac{2C_Q}{\sqrt{c_2(r_0)}}\log{\left(\sqrt{c_1(r_0)}\sqrt{c_2(r_0)}\right)}\\
&+\frac{2r_0^3}{a^2+Q^2}\frac{2C_Q}{\sqrt{c_2(r_0)}}\log{\left(c_2(r_0)+\sqrt{c_2(r_0)} \sqrt{c_1(r_0)+c_2(r_0)} \right)} \, .
\end{split}
\end{equation}
In the  SDL, by substituting (\ref{c_b_SDL}), $I_D(b)$ becomes
\begin{equation}
\begin{split}
I_D(b)\approx&-\frac{r_{sc}^3}{a^2+Q^2}\left(\frac{C_{-sc}}{\sqrt{c_{2 sc} z_{-sc}^2}}+\frac{C_{+sc} +C_{Qsc} z_{+sc}}{\sqrt{c_{2sc} z_{+sc}^2}}+\frac{C_{Qsc}}{\sqrt{c_{2sc}}} \right)\log{\left(\frac{b}{b_{sc}}-1\right)}\\
&+\frac{r_{sc}^3}{a^2+Q^2}\frac{C_{-sc}}{\sqrt{c_{2sc} z_{-sc}^2}}\log{\left[ \frac{16c_{2sc} ^2z_{-sc}^2 b_{sc}''}{c_{1sc}'^2 2 b_{sc}(z_{-sc}-1)^2}\right]}\\
&+\frac{r_{sc}^3}{a^2+Q^2}\frac{C_{+sc}+C_{Qsc} z_{+sc}}{\sqrt{c_{2} z_{+}^2}}\log{\left[ \frac{16 c_{2sc}^2 z_{+sc}^2 b_{sc}''}{c_{1}'^2 2b_{sc}(z_{+sc}-1)^2}\right]}
+\frac{r_{sc}^3}{a^2+Q^2}\frac{C_{Qsc}}{\sqrt{c_{2sc}}}\log{\left[ \frac{16 c_{2sc}^2 b_{sc}'')}{c_{1sc}'^2 2 b_{sc}}\right] } \, ,
\end{split}
\end{equation}
from which we can read off the coefficients $\bar a$ and $b_D$ as follows
\begin{equation}\label{abar_kn}
\begin{split}
\bar{a}=&\frac{r_{sc}^3}{\sqrt{c_{2sc}}}\left[ \frac{C_{-sc}}{r_{sc} r_--(a^2+Q^2)}+\frac{C_{+sc}}{r_{sc} r_{+}-(a^2+Q^2)}\right]
\end{split}
\end{equation}
%and
\begin{equation}\label{b_D_kn}
\begin{split}
b_D=&\bar{a} \log{\left[ \frac{8c_{2sc}^2 b_{sc}''}{c_{1sc}'^2 b_{sc}}\right]}\\
&+\frac{2r_{sc}^3}{\sqrt{c_{2sc}}} \left[ \frac{ C_{-sc}}{r_{sc} r_{-}-(a^2+Q^2)}\log{\left( 1-\frac{a^2+Q^2}{r_{sc} r_{-}}\right)}+ \frac{C_{+sc}+ C_{Qsc} z_{+sc} }{r_{sc} r_{+} -(a^2+Q^2)}\log{\left(1-\frac{a^2+Q^2}{r_{sc} r_{+}}\right)} \right]\, .
\end{split}
\end{equation}
They reduce to their counterparts in (\ref{abar_k}) and (\ref{bD}) respectively as $Q\to 0$.
As for the remaining contributions  to the regular part, and in the SDL, we have
\begin{equation}\label{b_R_kn}
\begin{split}
& b_R=I_R(r_{sc})=\int_0^1 f_R(z,r_{sc}) dz\\
&=\frac{2r_{0}^3}{a^2+Q^2}\frac{C_-}{\sqrt{c_2}z_-}\log{\left(\frac{z_-}{z_--1} \frac{2c_2+c_3+2\sqrt{c_2+c_3+c_4}\sqrt{c_2}}{4c_2} \right)}\\
&+\frac{2r_{0}^3}{a^2+Q^2}\frac{C_-}{z_-\sqrt{c_2+c_3 z_-+c_4z^2_-}}\log{\left( \frac{z_--1}{z_-}\frac{\left(\sqrt{c_2+c_3 z_-+c_4 z_-^2}+ \sqrt{c_2}\right)^2-c_4 z_-^2}{\left(\sqrt{c_2+c_3 z_-+c_4 z_-^2}+ \sqrt{c_2+c_3+c_4}\right)^2-c_4 (z_--1)^2}\right)}\\
&+\frac{2r_{0}^3}{a^2+Q^2}\frac{C_+}{\sqrt{c_2}z_+}\log{\left(\frac{z_+}{z_+-1} \frac{2c_2+c_3+2\sqrt{c_2+c_3+c_4}\sqrt{c_2}}{4c_2} \right)}\\
&+\frac{2r_{0}^3}{a^2+Q^2}\frac{C_++C_Qz_+}{z_+\sqrt{{c_2+c_3 z_-+c_4z^2_-}}}\log{\left( \frac{z_+-1}{z_-}\frac{\left(\sqrt{c_2+c_3 z_++c_4 z_+^2}+ \sqrt{c_2}\right)^2-c_4 z_+^2}{\left(\sqrt{c_2+c_3 z_++c_4 z_+^2}+ \sqrt{c_2+c_3+c_4}\right)^2-c_4 (z_+-1)^2}\right)}\\
&+\frac{2r_{0}^3}{a^2+Q^2}\frac{C_Q}{\sqrt{c_2}}\log{\left(\frac{z_+}{z_+-1}\right)}\Big\vert_{r_0=r_{sc}}
\end{split}
\end{equation}
In the limit of $Q\to 0$, where $C_Q$ and $c_4$ go to zero, the above expression of $b_R$ reduces to (\ref{bR_k}) in the Kerr case after implementing straightforward algebra. The coefficient $\bar{b}$  is  obtained using (\ref{b_D_kn}) and (\ref{b_R_kn}) as
%$
%\bar{b}=-\pi +b_D+b_R \,
%$
\begin{equation}\label{bbar_kn}
\begin{split}
\bar{b}=&-\pi+\bar{a} \log{\left[\frac{36}{4(1-c_{4sc}/c_{2sc})^2+4\sqrt{3}(1-c_{4sc}/c_{2sc})^{3/2}+3(1-c_{4sc}/c_{2sc})} \frac{8 c_{2sc}^2 b_{sc}''}{c_{1sc}'^2 b_{sc}}\right]}\\
&+\frac{r_{sc}^3}{\sqrt{c_{2sc}}}\frac{2(a^2+Q^2) C_{-sc}}{(a^2+Q^2-r_{sc}r_-)}\frac{\sqrt{3}}{{P_-}}\\
&\quad\quad \times \log\left[ \frac{-r_{sc}r_-}{a^2+Q^2-r_{sc}r_-}  \frac{\left({P_-}+\sqrt{3} ({a^2+Q^2}) \right)^2-3 \left({a^2+Q^2}-{r_{sc} r_-} \right)^2 ({c_{4sc}}/{c_{2sc}})}{\left(P_-+(a^2+Q^2) (1-c_{4sc}/c_{2sc})^{1/2}\right)^2-3 {r^2_{sc} r^2_-}  ({c_{4sc}}/{c_{2sc}})}\right]\\
&+\frac{r_{sc}^3}{\sqrt{c_{2sc}}}\frac{2 [(a^2+Q^2)C_{+sc} +(a^2+Q^2-r_{sc}r_+)C_{Qsc}]}{(a^2+Q^2-r_{sc}r_+)}\frac{\sqrt{3}}{{P_+}}\\
&\quad\quad \times \log{\left[ \frac{-r_{sc}r_+}{a^2+Q^2-r_{sc}r_+} \frac{\left({P_+}+\sqrt{3} ({a^2+Q^2}) \right)^2-3 \left({a^2+Q^2}-{r_{sc} r_+} \right)^2 ({c_{4sc}}/{c_{2sc}})}{\left(P_++(a^2+Q^2) (1-c_{4sc}/c_{2sc})^{1/2}\right)^2-3 {r^2_{sc} r^2_+}  ({c_{4sc}}/{c_{2sc}})}\right]} \, .
\end{split}
\end{equation}
In the equation above, we have replaced $c_{3sc} $ by the linear combination of $c_{2sc}$ and $c_{4sc}$ in (\ref{cs_kn}), given by
\begin{equation}
\begin{split}
c_{3sc}=&-\frac{2}{3}c_{2sc}-\frac{4}{3}c_{4sc} \, .
\end{split}
\end{equation}
We also have
\begin{equation}
\begin{split}
P^2_{\pm}&=(a^2+Q^2+2r_{sc}r_{\pm})(a^2+Q^2)-(a^2+Q^2+r_{sc}r_{\pm})(a^2+Q^2-r_{sc}r_{\pm})(c_{4sc}/c_{2sc})\, .
\end{split}
\end{equation}
Combining (\ref{br0_kn}),(\ref{rsc_kn}) and (\ref{bsc_kn}), the coefficients  $\bar a$ and $\bar b$ in (\ref{abar_kn}) and (\ref{bbar_kn}) can be analytically expressed as a function of the black hole's parameters in the SDL. Our results are ploted in Fig.(\ref{ab_kn_f}).
Again, notice that $\bar a>0$ but $\bar b<0$ with the parameters in the figure. Due to the fact that the bending angle for light rays resulting from the charged black hole is suppressed as compared with the neutral   black hole with the same impact parameter $b$, both $\bar a $ and $\vert \bar b \vert $ are found to increase with the charge $Q$.
%%%%%%%%%%%%%%%%%%%%%%%%%%%%%%%%%%%%%%%%%%%%%%%%%%%%%%%%%%%%%%%%%%%%%%%%%%%%%%%%%%%%%%%%%%%%%%%%%%%%%%%%%%%%%%%%%

\begin{figure}[h]
 \centering
 \includegraphics[width=0.63\columnwidth=0.63]{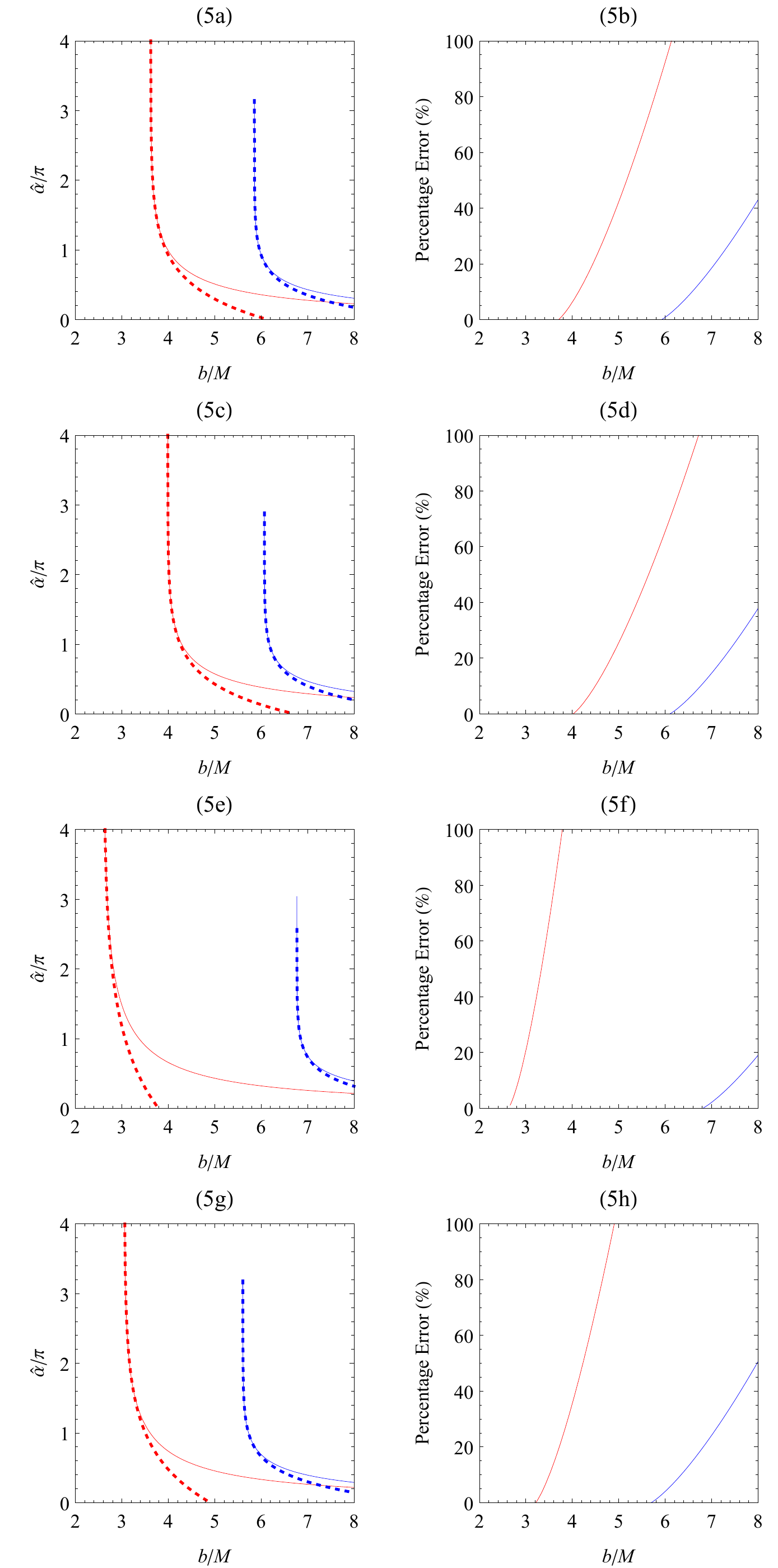}
 \caption{
%\textcolor{blue}
{ The SDL deflection angle (dotted lines) and the exact one (solid lines): (a)${a/M}=0.5$ and ${Q/M}=0.6$, (b) Error between them defined by $( \hat{\alpha}_{exact}-\hat{\alpha} )/ \hat{\alpha}_{exact} \times 100\%$; (c) ${a/M}=0.5$ and ${Q/M}=0.3$,(d) Error; (e) ${a/M}=0.9$ and ${Q/M}=0.3$, (f) Error; (g) ${a/M}=0.5$ and ${Q/M}=0.8$, (h) Error.}}
\label{alpha_com}
 \end{figure}
%%%%%%%%%%%%%%%%%%%%%%%%%%%%%%%%%%%%%%%%%%%%%%%%%%%%%%%%%%%%%%%%%%%%%%%%%%%%%%%%%%%%%%%%%%%%%%%%%%%%%%%%%%%%%%%%%

It is then quite straightforward to check that the coefficients  $\bar a$ and $\bar b$ in the Kerr-Newmann case can reduce to those in (\ref{abar_k}) and (\ref{bbar_k}) in the Kerr case by setting  $c_4 \to 0$ in the limit of $Q \to 0$, also leading to $P_{\pm} \to a \sqrt{a^2+2r_{sc}r_{\pm}}$.
To compare with the Reissner-Nordstr\"om black hole in \cite{Tsuka1,Tsuka2}, it is known that the impact parameter $b$ as a function of $r_0$ is
\begin{equation}\label{br0_Q}
b(r_0)=\frac{r_0^2}{\sqrt{Q^2-2Mr_0+r_0^2}}
\end{equation}
and the circular motion of light rays forms the photon sphere with the radius
\begin{equation}
r_{c}=\frac{3M+\sqrt{9M^2-8Q^2}}{2}\;.
\end{equation}
The critical impact parameter as a function of $r_{c}$ is given by
\begin{equation}
b_{c}=\frac{r_{c}^2}{\sqrt{Mr_{c}-Q^2}}\, .
\end{equation}
Notice the subscript is changed from $sc$ to $c$ since the same critical impact parameters are obtained for light rays in direct orbits and retrograde orbits in the case of the nonspinning black holes.
In the limit of  $a \to 0$, we have $C_{-sc} \to 0$ in (\ref{CpmQ_kn})  using the definition of $r_-$ in (\ref{rpm_kn}). Thus, the coefficient $\bar a$ in (\ref{abar_kn}) can be  further simplified using (\ref{CpmQ_kn}), (\ref{rpm_kn}) and $c_{1sc}=0$ giving
\begin{equation}
\bar{a}=\frac{r_{c}}{\sqrt{3M r_{c}-4 Q^2}}\, ,
\end{equation}
which reproduces the expression  in \cite{Tsuka1,Tsuka2}.
As for the coefficient $\bar b$, in the limit of $a\to 0$, apart from  $C_{-sc} \to 0$, $(a^2+Q^2)C_{+ sc} +(a^2+Q^2-r_{sc}r_+)C_{Qsc} \to 0$ as well.
So, the coefficient $\bar{b}$ in (\ref{bbar_kn}) has the contribution only from the term proportional to $\bar a$. After substituting  (\ref{br0_Q}) and (\ref{cs_kn}) in the  limit of $a\to 0$ to (\ref{bbar_kn}) and going through nontrivial algebra, we indeed recover the compact analytical expression  in \cite{Tsuka1,Tsuka2}:
\begin{equation}
\bar{b}= -\pi + \bar{a}\log{\left[ \frac{8(3Mr_{c}-4Q^2)^3}{M^2r_{c}^2(Mr_{c}-Q^2)^2} \left( 2\sqrt{Mr_{c}-Q^2}-\sqrt{3Mr_{c}-4Q^2} \right)^2  \right]}\, .
\end{equation}
Figure \ref{alpha_com} shows good agreement between the obtained SDL  expression  and the exact one in \cite{Hsiao} computed numerically when $b$ approaches $b_{c}$ for some values of $a$ and $Q$.

In conclusion, we have successfully achieved the analytical expression of the coefficient $\bar a $ and $\bar b$ in the form (\ref{hatalpha_as}) of the SDL deflection angle due to the spherically nonsymmetric black holes, although they look not as simple as  in the cases of  the spherically symmetric black holes.
Additionally, the obtained expressions can reproduce the respective ones due to the Kerr, Reissner-Nordstr\"om black holes and also due to the Schwarzschild black hole by taking the appropriate limits of the black hole's parameters.

\section{Relativistic Images of Gravitational lens and applications to supermassive galactic black holes}

We consider  the  cases of the planar light rays with the lens diagram shown in Fig.(\ref{fig:arch_02}), where $d_L$ and $d_S$ are the distances of the lens (black hole) and the light source from the observer, and also $d_{LS}$ represents the distance between the lens and  the source. The line joining the observer and the lens is considered as a reference optical axis.
The angular positions of the source and the image are measured from the optical axis, and are denoted by $\beta$ and $\theta$, respectively. The lens equation is given by
\begin{equation} \label{leneq}
%\tan\beta=\tan\theta- \frac{d_{LS}}{d_S}\tan\theta+\tan([\hat{\alpha}-2n\pi-\theta)]\;,
\tan s\beta=\tan\theta- \frac{d_{LS}}{d_S}[\tan\theta+\tan(\hat{\alpha}-\theta)]\;,
\end{equation}
where $\hat{\alpha}$ is the deflection angle of light rays obtained from (\ref{alpha_I}) that can be expressed in terms of the impact parameter $b$ as the light rays approach  to the black holes.
%
%\textcolor{blue}
{In \cite{Eiroa}, it is mentioned that the lens equations are applied for  the observer and the source immersed in the asymptotically flat spacetime, where the Kerr and Kerr-Newman black holes have the asymptotically flat metric.
Also, in the small $\beta$ and $\theta$ limits, we will see that the approximate lens equations to be found later are the same ones in \cite{Bozza_2003}, in which the Kerr black holes   are considered.}
In the SDL of our interest,  when the light rays wind around the black hole $n$ times, the deflection angle $\hat \alpha$  can be approximately by (\ref{hatalpha_as}). The  angle appearing in the lens equation should be within $2\pi$ and can be obtained from the deflection angle $\hat \alpha$ subtracting $ 2n\pi$.

Together with the relation between the impact parameter $b$ and the angular position of the image given by
\begin{equation} \label{b_theta}
b=d_L \sin\theta\;,
\end{equation}
in Fig.(\ref{fig:arch_02}), we can solve the lens equation (\ref{leneq}) with a given  angular position of the source  $\beta$ for  the angular position of the observed image $\theta$.
In the SDL,  when the angular position of the source is small,  $\theta$ is expectedly   small with the small impact parameter $b$. Then the lens equation (\ref{leneq}) can be further simplified by
\begin{equation}\label{leneq_app}
s\beta\simeq \theta-\frac{d_{LS}}{d_{S}}[\hat{\alpha}(\theta)-2n\pi] \,
\end{equation}
 and (\ref{b_theta}) can be approximated by  $b\simeq d_L\theta$.
% \textcolor{blue}
{This can reduce to the lens equations  in  \cite{Bozza_2003}, in which the small angle limits are considered.}
%the linear approximation of
% given by
%\begin{equation}
%\end{equation}
%
According to \cite{Bozza3}, the zeroth order solution $\theta_{sn}^0$ is obtained from $\hat{\alpha}(\theta_{sn}^0)=2n\pi$. Using the SDL deflection angle in (\ref{hatalpha_as}) we have then
\begin{equation}\label{theta0}
\theta_{sn}^0=\frac{\vert b_{sc}\vert}{d_L}\left( 1+e^{\frac{\bar{b}-2n\pi}{\bar{a}}} \right)
\end{equation}
for $n=1,2,\cdots$.
The angular position $\theta_{s n}$ decrease in $n$ and reaches the asymptotic angular position given by $\theta_{s\infty}= \vert b_{sc}\vert/{d_L}$ as $n\to \infty$.
With  the zeroth order solution (\ref{theta0}), the expansion of  $\hat{\alpha}(\theta)$ around $\theta=\theta_{sn}^0$ is written explicitly as
\begin{equation}
\hat{\alpha}(\theta)=\hat{\alpha}(\theta_{sn}^0)-\frac{ \bar a}{ e^{(\bar{b}-2n\pi)/\bar{a}}} \frac{d_L d_{LS} } {\vert b_{sc}\vert d_S} (\theta-\theta_{sn}^0)+{O}(\theta-\theta_{s n}^0)^2\, .
\end{equation}
Then  the approximate lens equation (\ref{leneq_app}) to the order $(\theta-\theta_{sn}^0)$ becomes
\begin{equation}
s\beta\simeq\theta_{sn}^0 +\left(1+\frac{ \bar a}{ e^{(\bar{b}-2n\pi)/\bar{a}}} \frac{d_L d_{LS}} {\vert b_{sc}\vert d_S}\right)(\theta-\theta_{sn}^0) \, .
\end{equation}
%
%by ignoring the first term in the parentheses as compared with the second term for $d_L \gg \vert b_{sc}\vert$,
Solving for $\theta$, by keeping  the lowest order term in $\vert b_{sc}\vert / d_L \ll 1 $, we find the angular position of the image as \cite{Bozza2}
\begin{equation}\label{theta_sn}
\begin{split}
\theta_{s n}\simeq\theta_{sn}^0+\frac{ e^{(\bar{b}-2n\pi)/\bar{a}}}{ \bar a} \frac{\vert b_{sc}\vert d_S}{d_{LS}d_L } (s \beta -\theta_{sn}^0) \;.
\end{split}
\end{equation}

We assume that either  Kerr or  Kerr-Newman black holes have the clockwise rotation shown in Fig.(\ref{fig:arch_02}).
The light rays emitted from the source circle around the black hole multiple times in the SDL  along a direct orbit (red line) with $s=+1$, where both the image and the source end up in the same sides of the optical axis with the angular position $\theta_{+n}$ and/or along a retrograde orbit (blue line) with $s=-1$, where the image and the source  are in the opposite sides with the angular position $\theta_{-n}$.
We also define  the angular position difference between the outermost image $\theta_{1\pm}$ and the asymptotic one near the optical axis as
\begin{equation}
\Delta \theta_{s}=\theta_{s1}-\theta_{s\infty} \,\, ,
\end{equation}
which is the value to compare with the resolution of the observation that allows to distinguish among a set of the relativistic images.
%
\begin{comment}
The magnifications of the images  $\mu$ defined to be the ratio of the flux of the image to the flux of the source can then be obtained from the ratio of the solid angles of the images and the sources, and are given  and further approximated in the SDL as
%
\begin{equation}
\mu=\left( \frac{\sin{\beta}}{\sin{\theta}}\frac{d  \beta }{d \theta} \right)^{-1} \approx \left( \frac{ \beta }{\theta}\frac{d  \beta}{d \theta} \right)^{-1}\, .
\end{equation}
%
For the image with the angular position $\theta_{sn}$, the magnification $\mu_{sn}$ is found to be \cite{Bozza2}
%
\begin{equation} \label{mu_sn}
\mu_{sn}\simeq \frac{ e^{(\bar{b}-2n\pi)/\bar{a}}}{ \bar a} \frac{\vert b_{sc}\vert d_S}{d_L d_{LS} }\frac{\theta^0_{sn}}{ \beta }\,.
\end{equation}
%
%
Apart from very small $\beta$, where the source and the lens are extremely aligned, $\mu$ normally is very small due to $\vert b_{sc}\vert \ll d_L$. It implies the demagnification of the images and thus causes difficulties in observing the relativistic images.
\end{comment}

 We now compute the angular positions  of the relativistic images of the sources for $n=1$ (the outermost image) due to either Kerr or Kerr-Newman black holes  with the mass $M=4.1\times 10^6 M_{\odot}$ and the distance $d_L=26000\;{\rm ly}$ of  the supermassive black hole Sagittarius A* at the center of our Galaxy as an example.
We also take  the ratio to be $d_{LS}/d_S=1/2$. In Table 1 (2), we consider both the image and the source are in the same (opposite)  sides of the optical axis, where the light rays travel along the direct (retrograde) orbits, and choose $\beta \sim \theta_{\pm 1}$.
The angular positions of the relativistic images are computed by (\ref{theta_sn}).  In the case of $\vert b_{sc}\vert  \ll d_L$, $\theta_{sn}$ is not sensitive to $\beta$ but mainly determined by $\theta_{sn}^0$ in (\ref{theta0}).
 Given $\bar a$ and $\vert \bar b\vert$ of the magnitudes shown in Fig.(\ref{ab_k_f}) and (\ref{ab_kn_f}), $e^{-\frac{\vert \bar{b}\vert+2n\pi}{\bar{a}}} \ll 1$. The behavior of $\theta_{sn}$ thus depends mainly on $\vert b_{sc}\vert$ as a function of angular momentum $a$ and  charge $Q$ of the black holes.

As discussed in the previous section, since the effects from the angular momentum of the black hole for direct orbits effectively induces more repulsive effects compared with the retrograde orbits, clearly shown in their effective potential $ W_\text{eff}$ (\ref{eq:Weff_k}),  the resulting $b_{+c} < \vert b_{-c}\vert $
%as illustrated in Fig.(\ref{bsc_a})
yields asymmetric values of $\theta_{+1} <\theta_{-1}$ for the same $a$ and $Q$. These features are shown in the Tables 1 and 2.
Additionally, we notice that $\theta_{+1}$ ($\theta_{-1}$) decreases (increase) in $a$ for fixed $Q$ resulting from  the decrease  (increase) of $ b_{+c}$ ($\vert b_{-c} \vert$) as $a$ increases.
As for $\Delta \theta$,   for the same $Q$,   $\Delta \theta_+$ increases with $a$ whereas $\Delta \theta_-$ decreases  with $a$.
In particular, $\Delta \theta_+$ can be increased from about $10^{-2} \mu\rm{as}$ with $a/M\sim 10^{-3}$ and $Q/M=10^{-3}$ to the value as high as $0.6 \mu{\rm as}$ with $a/M =0.9$ and $Q/M=10^{-3}$, which certainly increases their observability by the current very long baseline interferometry (VLBI) \cite{Ulv,Johnson_2020}.
As for the finite $Q$ effects, also showing the repulsion to the light rays seen in the effective potential (\ref{eq:Weff_kn}), both $\theta_{+1}$ and $\theta_{-1}$ decrease in $Q$ for fixed $a$, resulting in the slightly increase of $\Delta \theta_{\pm}$ as $Q$ increases.

\begin{figure}[h]
 \centering
 \includegraphics[width=0.88\columnwidth=0.88]{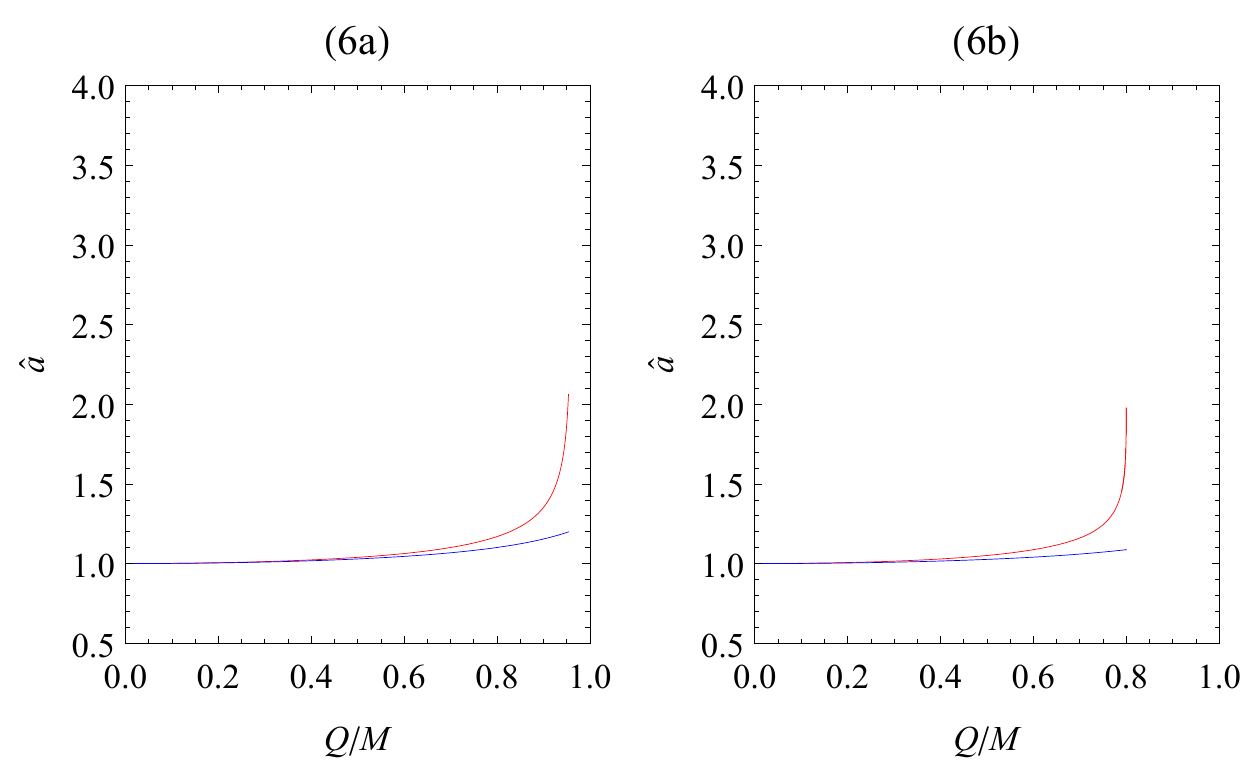}
 \caption{
%\textcolor{blue}
{
%\textcolor{blue}
{The coefficient  $\hat a$ as a function of the black hole charge $Q/M$ for the direct (retrograde) orbits with (a) $a/M=0.3$, (b) $a/M=0.6$.  The display of the plot follows the convention in Fig.(\ref{bsc_a}).}}}
\label{Fig6}
\end{figure}

\begin{table}
\begin{tabular}{ccccccc}
\hline
\hline
 $a/{M}$  ~  & $Q/{M}$ ~   &$\theta_{+1}$ ($\mu$as)      &$\hat{\alpha}$         &$b/M$          &$\theta_{+\infty}$ ($\mu$as)  &$\Delta \theta_{+}$ ($\mu$as)\\
\hline
$10^{\tiny -3}$ & $10^{\tiny -3}$    &26.4231      & $2\pi+32.8135$ ($\mu$as)     & $5.2007$          &26.3900   &0.0331\\
$ $ & $0.3$                                       &26.0217     & $2\pi+32.0563$ ($\mu$as)     & $5.1217$           &25.9866   &0.0351\\
$ $ & $0.6$                                       &24.7179     & $2\pi+29.4336$ ($\mu$as)     & $4.8651$           &24.6747   &0.0432\\
$ $ & $0.8$                                       &23.1445     & $2\pi+26.2837$ ($\mu$as)     & $4.5554$           &23.0849   &0.0596\\
\hline
$0.5$ & $10^{\tiny -3}$                   &20.9290      & $2\pi+21.8561$ ($\mu$as)     & $4.1193$          &20.8119   &0.1171\\
$ $ & $0.3$                                      &20.4085     & $2\pi+20.8203$ ($\mu$as)     & $4.0169$           &20.2758   &0.1327\\
$ $ & $0.6$                                      &18.6189     & $2\pi+17.2398$ ($\mu$as)     & $3.6646$           &18.4049   &0.2140\\
$ $ & $0.8$                                      &16.0922     & $2\pi+12.1835$ ($\mu$as)     & $3.1673$           &15.5372   &0.5550\\
\hline
$0.9$ & $10^{\tiny -3}$                  &15.1170      & $2\pi+10.2354$ ($\mu$as)     & $2.9754$          &14.4517   &0.6653\\
$ $ & $0.3$                                     &14.1818      & $2\pi+8.36638$ ($\mu$as)     & $2.7913$          &13.2701   &0.9117\\
$ $ & $0.6$                                     &$\cdots$                  &$\cdots$                                                 &$\cdots$                         &$\cdots$              &$\cdots$          \\
$ $ & $0.8$                                     &$\cdots$                     &$\cdots$                                                &$\cdots$                            &$\cdots$               &$\cdots$           \\
\hline
\hline
\end{tabular}
\caption{\label{tab:table-I}Relativistic images on the same side of the source with the angular position $\beta=10$ ($\mu$as) where the light rays are along direct orbits seen in Fig.(\ref{fig:arch_02}).}
\end{table}
\begin{table}
\begin{tabular}{ccccccc}
\hline
\hline
 $a/{M}$ ~    & $Q/{M}$ ~   &$\theta_{-1}$ ($\mu$as)      &$\hat{\alpha}$         &$b/M$              &$\theta_{-\infty}$ ($\mu$as) & $\Delta \theta_- $ ($\mu$as)\\
\hline
$10^{\tiny -3}$ & $10^{\tiny -3}$    &26.4433      & $2\pi+72.8286$ ($\mu$as)     & $5.20464$          &26.4103   &0.0330\\
$ $ & $0.3$                                       &26.0422     & $2\pi+72.0654$ ($\mu$as)     & $5.12569$           &26.0073   &0.0349\\
$ $ & $0.6$                                       &24.7395     & $2\pi+69.4844$ ($\mu$as)     & $4.86931$           &24.6966   &0.0429\\
$ $ & $0.8$                                       &23.1680     & $2\pi+66.3165$ ($\mu$as)     & $4.56000$           &23.1088   &0.0592\\
\hline
$0.5$ & $10^{\tiny -3}$                   &31.1994      & $2\pi+82.4458$ ($\mu$as)     & $6.14075$          &31.1862    &0.0132\\
$ $ & $0.3$                                      &30.8561      & $2\pi+81.6482$ ($\mu$as)     & $6.07318$           &30.8422   &0.0139\\
$ $ & $0.6$                                      &29.7638      & $2\pi+79.5352$ ($\mu$as)     & $5.85820$           &29.7479   &0.0159\\
$ $ & $0.8$                                      &28.5058      & $2\pi+76.9916$ ($\mu$as)     & $5.61059$           &28.4866   &0.0192\\
\hline
$0.9$ & $10^{\tiny -3}$                  &34.7203       & $2\pi+89.4397 $($\mu$as)     & $6.83374 $          &34.7130   &0.0073\\
$ $ & $0.3$                                     &34.4063       & $2\pi+88.5984$ ($\mu$as)     & $6.77195 $          &34.3988   &0.0075\\
$ $ & $0.6$                                     &$\cdots$                  & $\cdots$                                              &$\cdots$                            &$\cdots$               &$\cdots$          \\
$ $ & $0.8$   &$\cdots$                  &$\cdots$    &$\cdots$          &$\cdots$    &$\cdots$          \\
\hline
\hline
\end{tabular}
\caption{\label{tab:table-II}Relativistic images on the opposite side of the source with the angular position $\beta=10$ ($\mu$as) where the light rays are in retrograde orbits seen in Fig.(\ref{fig:arch_02}).}
\end{table}
%
%
%\textcolor{blue}
{Another application of the analytical expression of the deflection angle   on the equatorial plane is to consider the quasiequatorial gravitational lensing based upon  the works of \cite{Bozza_2003,Gyu_2007}.  In this situation,  the polar angle $\theta$ is  set to be slightly away from $\theta=\frac{\pi}{2}$ and now becomes time dependent.
In the SDL,  the deflection angle of light rays with the additional initial declination can also be cast into the form of (\ref{hatalpha_as}) where the coefficients are replaced by $\hat a$ and $\hat b$.
In particular, the coefficient $\hat a$ obtained from the slightly off the equatorial plane can be related by the coefficient $\bar a$ on the equatorial plane through the $\omega$ function as
\begin{equation} \label{hata}
\hat a=\omega (r_{sc}) \, \bar a\;,
\end{equation}
where $\omega$ depends on $r$, and in turn depends on the deflection angle $\phi(r)$.
Notice that the above relation (\ref{hata}) involves $\omega$, which is  evaluated  at $r=r_{sc}$. In the case of the Kerr black hole, it is found that \cite{Bozza_2003} }
\begin{equation} \label{omega_k}
\omega(r_{sc})=\frac{(r_{sc}^2+a^2-2M r_{sc}) \sqrt{b_{sc}^2-a^2}}{2M a r_{sc}+b_{sc}(r_{sc}^2-2 M r_{sc})}\, ,
\end{equation}
%
%\textcolor{blue}
{and thus for the Schwarzschild case we have $a\rightarrow 0$, $\omega \rightarrow 1$.
Then, substituting (\ref{rsc_k}) and (\ref{bsc_k}) into (\ref{omega_k}), together with  the expression of $\bar a$ in (\ref{abar_k}), through (\ref{hata}) gives $\hat a=1$ for the Kerr case.
However, in the Kerr-Newman black hole, the straightforward calculations show that the above relation (\ref{hata}) still holds true.
 The detailed derivations will appear in our future publication. Thus, the coefficient $\hat a$ can be analytically given by the coefficient $\bar a$ in (\ref{abar_kn}), together with the $\omega$ function in the Kerr-Newman case in below
\begin{equation}\label{omega_kn}
\omega (r_{sc})=\frac{(r_{sc}^2+a^2+Q^2-2 M r_{sc}) \sqrt{b_{sc}^2-a^2}}{- a (Q^2-2M  r_{sc})+b_{sc}(r_{sc}^2+Q^2-2 M r_{sc})}\, .
\end{equation}
As $Q\rightarrow 0$, (\ref{omega_kn}) reduces to (\ref{omega_k}) in the Kerr case.
The behavior of $\hat a$  as a function of the charge $Q$ with the choices of the angular momentum $a$ for direct and retrograde orbits is  displayed in Fig.(\ref{Fig6}). The value of $\hat a$ ($\hat a>1$) increases with $Q$ for both direct and retrograde orbits.
According to \cite{Bozza_2003,Gyu_2007},  the magnification of relativistic  images might formally diverge when the angular positions of the sources are at caustic points.  The corresponding magnifying power close to caustic points due to the light rays winding around the black hole $n$ times is given by $\bar \mu_n$ with the ratio between two neighboring caustic points
 \begin{equation}
 \frac{\bar \mu_{n+1}}{\bar \mu_n} \propto e^{-\pi/{\hat a}} \,
 \end{equation}
depending only on $\hat a$.
In the Kerr case with $\hat a=1$, this ratio is independent of the black hole angular momentum $a$, whereas in the Kerr-Newman case with $\hat a >1$ shown in the Fig.(\ref{Fig6}), the ratio decreases with $Q$ for both direct and retrograde orbits \cite{Gyu_2007}.
Here we just sketch some of the effects from the charge $Q$ of the black hole on the magnification of  relativistic images. To have the full pictures of the caustic points and find the magnification of relativistic images, it in fact deserves the extensive study to compute not only $\hat a$ but also $\hat b$ by following \cite{Bozza_2003,Gyu_2007}.  The further  extension from quasiequatorial plane to the full sky is also of great interest \cite{Gralla_2020a,Gralla_2020b,Johnson_2020}. }
 %
 %======================================================================================================================
 %======================================================================================================================

%\textcolor{blue}
\begin{comment}
{The magnifications of the relativistic images at the angular position $\theta_{sn}$ can be calculated from (\ref{mu_sn}), and their dependence of the black hole's parameters is determined by the coefficients $\bar a$ and $\bar b$,  mainly through $\mu \propto e^{-2n\pi/\bar a}$ since typically $\vert \bar b \vert <2\pi$.
%
In the case of $n=1$, the increase (decrease) of $\bar a$ for direct (retrograde) orbits as  $a$ increases by fixing $Q$ drives the increase (decrease) in the magnification $\mu_{+1}$ ($\mu_{-1}$), as shown in Tables 1 and 2 respectively.
%
In particular, $\mu_{+1}$ for direct orbits can be increased from $10^{-13}$ with  $a/M=10^{-3}$ and $Q/M=10^{-3}$ to $10^{-12}$ with  $a/M =0.9 $ and $Q/M =10^{-3}$.
%
Similar discussions are also applied to the charge $Q$ of the black hole with  fixed angular momentum $a$.  The increase of $\bar a$ as $Q$ increases for both direct and retrograde orbits results in the increase of $\mu_{\pm}$ also shown in the Tables.
%
For example, $\mu_{\pm1}$  can also be increased from $10^{-13}$ with  $a/M=10^{-3}$ and $Q/M=10^{-3}$ to $10^{-12}$ with  $a/M =10^{-3}$ and  $ Q/M=0.8$.  The effects from angular momentum $a$ and charge $Q$ of the black holes with the enhanced $\mu$ values may  increase the visibility of the created images.}
\end{comment}

\section{Summary and outlook}
In summary, the dynamics of light rays traveling around the Kerr black hole and the Kerr-Newman black hole, respectively, is studied with the detailed derivations on achieving analytical expressions of $\bar a$ and $\bar b$ in the approximate form of the deflection angle in the SDL.
Various known results are checked by taking the proper limits of the black hole's parameters. The analytical expressions are then applied to compute the angular positions  of relativistic images due to the supermassive  galactic black holes.
We find that the effects from the angular momentum $a$ for direct orbits of light rays  and the charge $Q$  for both direct and retrograde orbits increase the angular separation of the outermost images from the others.
Although the observation of relativistic images is a very difficult task \cite{Ulv}, our studies show potentially increasing observability of the relativistic images from the effects of angular momentum and charge of the black holes.
Hopefully, relativistic images will be observed in the near future. Through the analytical results we present in this work, one can reconstruct the black hole's parameters that give strong lensing effects.
%
%\textcolor{blue}
{As light rays travel on the quasiequatorial plane, our analytical results on the equatorial plane can also be applied to roughly estimate the relative  magnifications of relativistic images with the sources   near one of the caustic points by taking account of  the dynamics of the light rays in the polar angle. The work of investigating  the structure of the caustic points from the effects of the charge $Q$ of the Kerr-Newman black holes and  the magnification of relativistic images  is in progress.}
Also, inspired by the recent advent of horizon-scale observations of astrophysical black holes, the properties of null geodesics become of great relevance to astronomy.
The recent work of \cite{Gralla_2020a,Gralla_2020b,Johnson_2020}  provides an extensive analysis on Kerr black holes. We also plan to extend the analysis of null geodesic to Kerr-Newman black holes, focusing on the effects from the charge of black holes.

\begin{acknowledgments}
This work was supported in part by the Ministry of Science and Technology, Taiwan, under Grant No.109-2112-M-259-003.
\end{acknowledgments}

\end{document}